\documentclass[11pt]{article}
\usepackage[preprint]{acl}
\usepackage{caption}
\usepackage{times}
\usepackage{latexsym}
\usepackage[T1]{fontenc}
\usepackage[utf8]{inputenc}
\usepackage{microtype}
\usepackage{inconsolata}
\usepackage{float}
\usepackage{graphicx}
\graphicspath{{figures/}}
\usepackage{booktabs}
\usepackage{natbib}
\usepackage{multirow}
\usepackage{amsmath}
\usepackage{amssymb}
\usepackage{enumitem}
\usepackage{url}
\usepackage{dblfloatfix}
\usepackage{tcolorbox}
\usepackage{algorithm}
\usepackage{algpseudocode}
\algrenewcommand\algorithmicrequire{\textbf{Input:}}
\algrenewcommand\algorithmicensure{\textbf{Output:}}

\title{GuardRAG: A Post-Retrieval Defense against Poisoning Attacks in Retrieval-Augmented Generation}

\author{
    Linghui Jiao\textsuperscript{1},
    Kang Wei\textsuperscript{1},
    Fushuo Huo\textsuperscript{1},
    Ruijie Zhao\textsuperscript{1}, \\
    Yiyang Ni\textsuperscript{2},
    Jingcai Guo\textsuperscript{3},
    Yipeng Zhou\textsuperscript{4},
    Jun Li\textsuperscript{1} \\[0.5em]
    \textsuperscript{1}Southeast University, China \\
    \textsuperscript{2}Jiangsu Second Normal University, China \\
    \textsuperscript{3}The Hong Kong Polytechnic University, Hong Kong SAR, China \\
    \textsuperscript{4}Macquarie University, Australia \\[0.3em]
    \texttt{jiaolinghui@seu.edu.cn}
}

\begin{document}
\maketitle

\begin{abstract}

Retrieval-Augmented Generation (RAG) enhances the factual grounding of Large Language Models (LLMs) by incorporating external evidence, but also exposes them to corpus poisoning attacks. In particular, coordinated attacks can introduce multiple mutually reinforcing poisoned documents into retrieved contexts, making reliable evidence filtering challenging. To address this problem, we propose GuardRAG, a training-free and plug-and-play post-retrieval defense that exploits the semantic structure of retrieved evidence to identify suspicious document groups. Specifically, GuardRAG first expands the retrieval range and removes near-duplicate candidates, then partitions the retained documents using Hierarchical Agglomerative Clustering (HAC) with a query-adaptive threshold. Finally, it scores and prunes suspicious clusters based on query relevance, cluster size, and semantic compactness. Experiments across three question-answering datasets, multiple poisoning attacks, and six LLMs demonstrate that GuardRAG substantially reduces attack success rates while maintaining answer utility under clean retrieval.

\end{abstract}

\section{Introduction}
Large Language Models (LLMs) have demonstrated remarkable performance across a wide range of knowledge-intensive tasks, including open-domain question answering and multi-step reasoning. However, hallucination issues and the limitations of static parametric knowledge fundamentally constrain their practical capabilities~\cite{huang2025survey,sahoo2024comprehensive}. Retrieval-Augmented Generation (RAG) addresses these limitations by incorporating external corpora, enabling LLMs to access supporting evidence beyond their parametric memory~\cite{lewis2020retrieval}.

However, reliance on external corpora exposes RAG systems to security threats, particularly corpus poisoning attacks that manipulate retrieved evidence to influence downstream generation~\cite{zou2025poisonedrag,su2025corpus,zhang2025benchmark,xue2024badrag,cheng2024trojanrag}. In such attacks, adversaries inject malicious passages into external knowledge sources and optimize their relevance to target queries, increasing the likelihood that poisoned evidence will be retrieved. Since RAG systems often rely on large-scale or continuously updated corpora, verifying the trustworthiness of every document before retrieval is impractical. This motivates defenses that identify and filter poisoned evidence after retrieval but before generation.

\begin{figure*}[t]
    \centering
    \includegraphics[width=\textwidth]{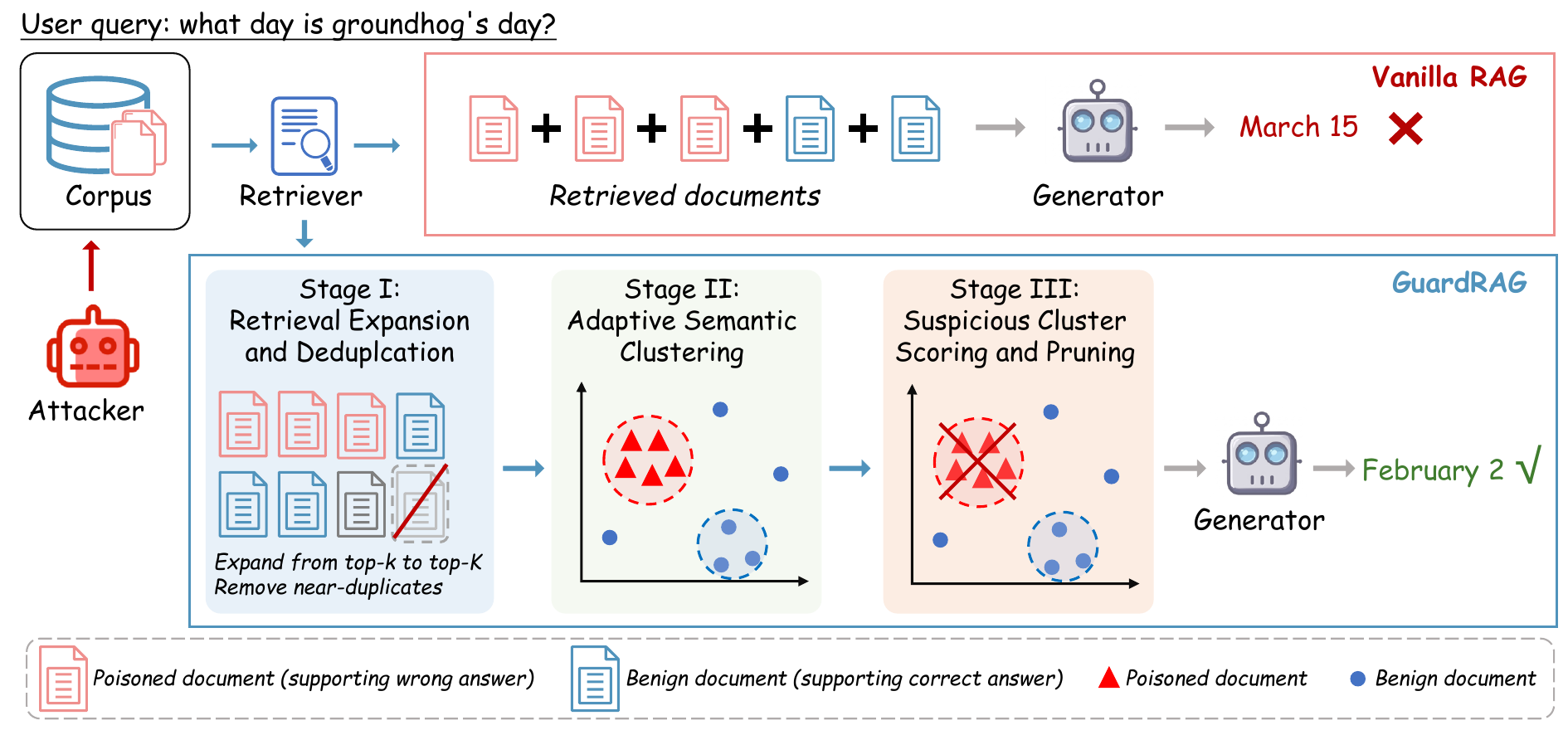}
    \caption{Comparison between vanilla RAG and GuardRAG under corpus poisoning attacks. While vanilla RAG can be misled by poisoned retrieved documents, GuardRAG mitigates such attacks through retrieval expansion and deduplication, adaptive semantic clustering, and suspicious cluster pruning before generation.}
    \label{fig:bigfigure}
\end{figure*}

To counter corpus poisoning, existing RAG defenses employ strategies such as robust aggregation~\cite{xiang2024robustrag}, reliability-aware scoring~\cite{shen2025reliabilityrag}, knowledge consolidation~\cite{wang2025astute}, and self-reflective generation~\cite{asai2023selfrag}. However, high-intensity coordinated poisoning poses a particular challenge: multiple poisoned documents can reinforce the same attacker-specified misinformation and dominate the retrieved evidence, making conventional evidence aggregation and reliability assessment less effective. This motivates examining the collective structure of retrieved documents rather than assessing each passage independently.

To address this challenge, we introduce GuardRAG, a training-free and plug-and-play post-retrieval defense framework designed to defend RAG systems against coordinated corpus poisoning attacks. As illustrated in Figure~\ref{fig:bigfigure}, while standard RAG is easily misled by poisoned documents, GuardRAG mitigates attacks by shifting the focus from individual document assessment to group-level pattern detection. Specifically, our framework follows a three-stage pipeline: it first expands the retrieval range and removes near-duplicate candidates, then employs adaptive semantic clustering to group semantically consistent documents, and finally prunes suspicious clusters prior to the generation phase.

Our contributions can be summarized as follows:
\begin{itemize}
    \item We identify a characteristic structural pattern of coordinated corpus poisoning: retrieved poisoned documents tend to exhibit stronger local semantic concentration than benign evidence in the embedding space.
    \item We propose GuardRAG, a training-free, plug-and-play post-retrieval defense that filters coordinated poisoned evidence through three stages: retrieval expansion with deduplication, adaptive semantic clustering, and suspicious-cluster pruning.
    \item We evaluate GuardRAG across three datasets, multiple retrievers, six generators, and several poisoning strategies, demonstrating substantially lower attack success while preserving answer utility under standard clean retrieval.
\end{itemize}

\section{Related Work}

\textbf{Retrieval-Augmented Generation.} RAG typically consists of two core components: a \emph{retriever} and a \emph{generator}~\cite{lewis2020retrieval}. The retriever retrieves documents relevant to a user query from an external corpus, while the generator produces the final answer conditioned on the retrieved evidence. Retrieval methods can generally be divided into sparse and dense retrieval. Sparse retrieval relies on lexical matching, with representative methods including TF-IDF~\cite{salton1988term} and BM25~\cite{robertson1994okapi}. Dense retrieval instead performs retrieval based on embedding similarity in continuous vector spaces, with representative approaches including DPR~\cite{karpukhin2020dense}, ANCE~\cite{xiong2020ance}, and Contriever~\cite{izacard2021contriever}.

\textbf{Corpus Poisoning Attacks.} Despite its effectiveness, RAG also introduces serious security vulnerabilities, particularly corpus poisoning attacks. \emph{PoisonedRAG} demonstrates that malicious documents injected into external corpora can manipulate downstream generation results~\cite{zou2025poisonedrag}. Subsequent studies, including \emph{POISONCRAFT}~\cite{shao2025poisoncraft}, \emph{AgentPoison}~\cite{chen2024agentpoison}, \emph{BadRAG}~\cite{xue2024badrag}, \emph{CPA-RAG}~\cite{li2025cpa}, and \emph{PR-Attack}~\cite{jiao2025pr}, further investigate targeted poisoning attacks against specific queries or outputs. Among them, \emph{POISONCRAFT} improves attack effectiveness through adversarial optimization of poisoned documents; \emph{AgentPoison} and \emph{BadRAG} combine poisoned knowledge with backdoor-trigger mechanisms; and \emph{PR-Attack} enhances retrieval success through collaborative optimization across multiple poisoned documents.

\textbf{Defense Methods.} Existing defense methods can be broadly categorized into three groups. The first group, including \emph{RobustRAG}~\cite{xiang2024robustrag}, \emph{InstructRAG}~\cite{wei2024instructrag}, and \emph{ReliabilityRAG}~\cite{shen2025reliabilityrag}, improves robustness by aggregating or assessing retrieved evidence, often introducing additional verification, reasoning, or token-processing overhead. The second group, represented by \emph{ASTUTE RAG}~\cite{wang2025astute}, \emph{TrustRAG}~\cite{zhou2025trustrag}, and \emph{Self-RAG}~\cite{asai2023selfrag}, leverages self-reflection, internal knowledge, or iterative reasoning to resolve unreliable or conflicting retrieved evidence, generally requiring additional inference-time computation. The third group, including \emph{Why So Gullible?}~\cite{hong2024gullible}, \emph{RECOMP}~\cite{xu2023recomp}, and \emph{Learning to Filter}~\cite{wang2023learning}, improves robustness through learned filtering, compression, or context-selection mechanisms, which require auxiliary training or task-specific data.

Compared with these approaches, GuardRAG shifts the defense perspective from individual-document assessment to group-level structural analysis of retrieved evidence. It operates after retrieval and before generation, and requires neither auxiliary model training nor additional LLM generation calls, making it compatible with RAG pipelines as a training-free and plug-and-play defense.

\section{Problem Formulation}
\label{sec:problem_setup}

In this section, we formalize the RAG process and define the threat model in this work according to the attacker's goal and capability, following prior studies on corpus poisoning attacks against RAG systems \cite{zou2025poisonedrag,li2025cpa}.

\subsection{Task Formulation}
A standard RAG system comprises a knowledge corpus $\mathcal{C}$, a retriever $\mathcal{R}$, and a generator $\mathcal{G}$. Given a user query $q$, the retriever fetches a subset of top-$k$ relevant documents $\mathcal{D}_q = \mathcal{R}(q, \mathcal{C})$. The generator then conditions on both the query and the retrieved context to produce the answer $y = \mathcal{G}(q, \mathcal{D}_q)$. We aim to safeguard the integrity of $y$ even when the corpus $\mathcal{C}$ is compromised by malicious poisoning.

\subsection{Threat Model}
We focus on a knowledge-misleading scenario, where an attacker aims to manipulate the RAG system into generating incorrect answers that convey specific misinformation for a target query $q^*$.

\paragraph{Attacker's Goal.}
The attacker's objective is to induce the system into generating a response $y$ that contains a specific, pre-determined target phrase $y^*$ as a substring. Following established protocols \cite{zou2025poisonedrag}, an attack is deemed successful if the target phrase $y^*$ is present within the generated output, where $y^*$ typically represents a short phrase or entity of misinformation.

\paragraph{Attacker's Capability.}
The attacker can inject a limited budget of $b$ poisoned passages, denoted by $\mathcal{M}$, into the corpus $\mathcal{C}$, with $b=|\mathcal{M}|\ll|\mathcal{C}|$. The attacker crafts these passages to increase their retrieval likelihood for a target query $q^*$ and reinforce the same attacker-specified incorrect answer through mutually supporting evidence.

\paragraph{Attacker's Knowledge.}
Referring to the setup of PoisonedRAG, we assume that attackers cannot obtain the internal parameters or weights of $\mathcal{G}$, yet they are able to access specific user queries and detailed parameters of $\mathcal{R}$.

\section{GuardRAG}
\subsection{Motivation}
\begin{figure}[t]
    \centering
    \includegraphics[width=0.98\columnwidth]{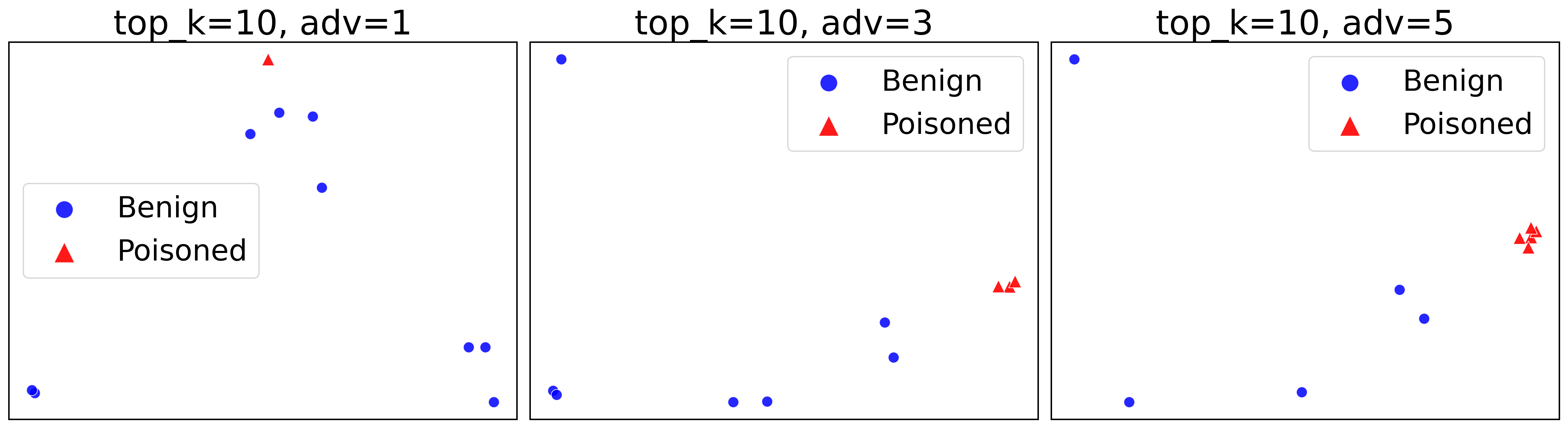}
    \caption{Embedding-space distributions of retrieved documents under different poisoning intensities. \textcolor{red}{Poisoned documents} (red triangles) exhibit stronger local concentration than \textcolor{blue}{benign documents} (blue circles).}
    \vspace{-8pt}
    \label{fig:retrieval_distribution}
\end{figure}

The design of GuardRAG is motivated by the distinct structural patterns of retrieved evidence in the embedding space under adversarial conditions. We use $\mathrm{adv}_k$ to denote an attack setting with $k$ poisoned documents in the retrieved context. As shown in Figure~\ref{fig:retrieval_distribution}, coordinated poisoned documents tend to exhibit stronger local semantic concentration than benign evidence, with the distinction becoming more pronounced as the number of retrieved poisoned documents increases.

To further validate this observation, we analyze retrieved documents across three attacks and three datasets. Among 900 query-level instances, poisoned documents exhibit smaller within-group distances than poisoned-to-benign distances in 896 cases. Moreover, all retrieved poisoned documents are assigned to a single HAC cluster in 862 cases, indicating that coordinated poisoning frequently produces identifiable semantic groups. Detailed structural statistics are provided in Appendix~\ref{sec:structural_validation}.

Building on this observation, GuardRAG employs a three-stage pipeline: retrieval expansion and deduplication, adaptive semantic clustering, and suspicious cluster pruning. Algorithm~\ref{alg:guardrag} summarizes the procedure, and the following subsections describe each stage.

\begin{algorithm}[H]
\small
\caption{GuardRAG}
\label{alg:guardrag}
\begin{algorithmic}[1]
\Require query $q$; corpus $\mathcal{C}$; retriever $\mathcal{R}$; generator $\mathcal{G}$; encoder $E$; retrieval depth $k$; expansion factor $m$; deduplication threshold $\gamma$; threshold cap $\tau_{\max}$
\Ensure final answer $y$

\Statex \texttt{// Stage 1: Retrieval Expansion and Deduplication}
\State $K \leftarrow mk$
\State $\widetilde{R}_q \leftarrow \mathcal{R}(q,\mathcal{C},K)$
\State $R_q \leftarrow \operatorname{Deduplicate}(\widetilde{R}_q,E,\gamma)$
\State $N_q \leftarrow |R_q|$

\Statex \texttt{// Stage 2: Adaptive Semantic Clustering}
\State $\mathbf{q}, \{\mathbf{h}_j\}_{j=1}^{N_q} \leftarrow E(q,R_q)$
\State $\hat{\tau}_q \leftarrow \operatorname{AdaptiveThreshold}(\{\mathbf{h}_j\}_{j=1}^{N_q})$
\State $\tau_q \leftarrow \hat{\tau}_q$ if $\hat{\tau}_q \leq \tau_{\max}$ else $0$
\State $\mathcal{C}_q \leftarrow \operatorname{HAC}(\{\mathbf{h}_j\}_{j=1}^{N_q},\tau_q)$

\Statex \texttt{// Stage 3: Suspicious Cluster Pruning}
\State $\mathcal{I}_q \leftarrow \{C_i \in \mathcal{C}_q \mid |C_i| \geq 3\}$
\If{$\mathcal{I}_q \neq \emptyset$}
    \State $C^* \leftarrow \arg\max_{C_i \in \mathcal{I}_q} S_{\mathrm{susp}}^{(i)}$
    \State $R_q' \leftarrow R_q \setminus C^*$
\Else
    \State $R_q' \leftarrow R_q$
\EndIf

\Statex \texttt{// Final Answer Generation}
\State $y \leftarrow \mathcal{G}(q,R_q')$
\State \Return $y$
\end{algorithmic}
\end{algorithm}

\subsection{Stage I: Retrieval Expansion and Deduplication}

Given a query $q$, standard RAG retrieves the top-$k$ documents, whereas GuardRAG first expands the retrieval range to the top-$K$ candidates, where $K = mk$ and $m > 1$:
\begin{equation}
\widetilde{R}_q = \{d_1, d_2, \dots, d_K\}.
\end{equation}

The expanded retrieval range introduces additional relevant evidence for subsequent structural analysis. To reduce redundancy, GuardRAG further removes near-duplicate documents according to their embedding similarity. Documents with cosine similarity greater than or equal to a threshold $\gamma$ are treated as near-duplicates, and only the higher-ranked document is retained. In our experiments, we set $\gamma = 0.98$. The resulting document set is denoted as $R_q$, with $N_q = |R_q| \leq K$.

\subsection{Stage II: Adaptive Semantic Clustering}

To reveal the semantic structure of the expanded retrieval set, GuardRAG encodes the query $q$ and all documents in $R_q$ using \texttt{bge-large-en-v1.5} \cite{bge_embedding}. All embeddings are L2-normalized before distance computation. We denote the normalized query embedding as $\mathbf{q} \in \mathbb{R}^d$ and the normalized embedding of each document $d_j \in R_q$ as $\mathbf{h}_j \in \mathbb{R}^d$, where $d$ is the embedding dimension.

\begin{figure*}[h]
    \centering
    \includegraphics[width=0.98\textwidth]{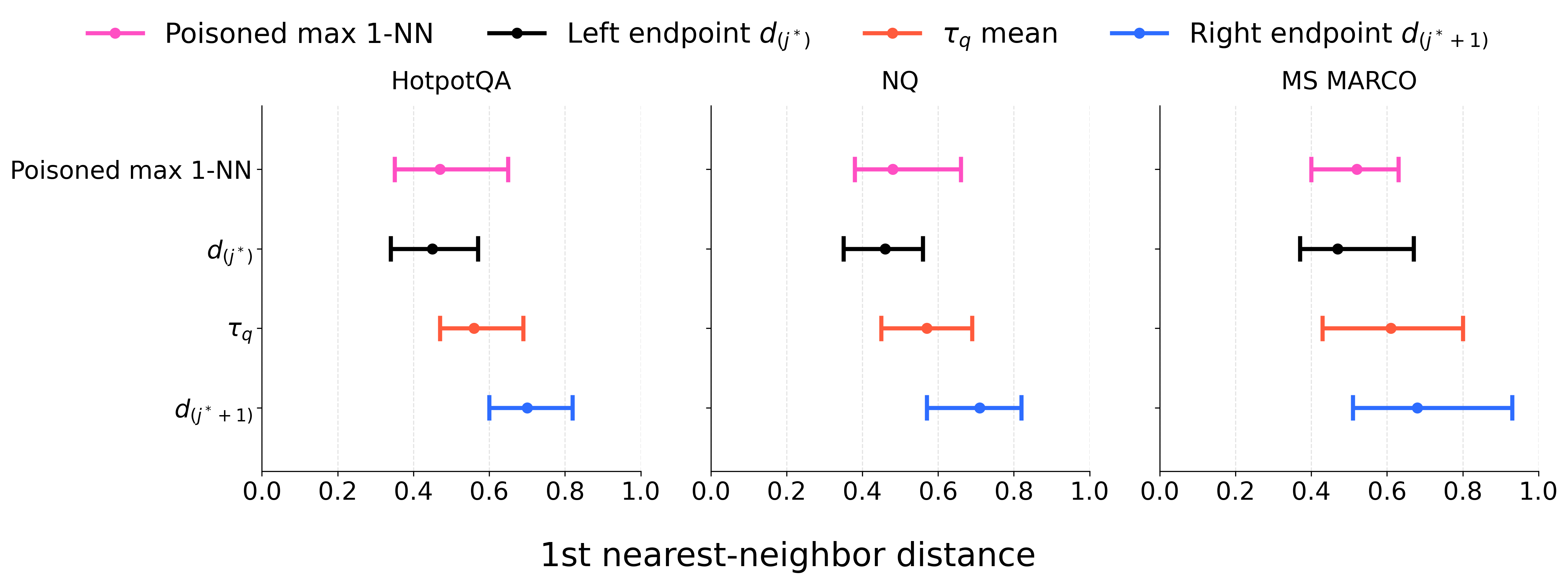}
    \caption{Empirical validation of the adaptive threshold $\tau_q$ across Natural Questions (NQ)~\cite{kwiatkowski2019natural}, HotpotQA~\cite{yang2018hotpotqa}, and MS MARCO~\cite{bajaj2016msmarco}. We show the maximum poisoned 1-NN distance, the two endpoints of the largest adjacent gap in the sorted 1-NN distances, and the resulting threshold. Error bars indicate empirical percentile ranges.}
    \vspace{-8pt}
    \label{fig:threshold_validation}
\end{figure*}

Based on the structural pattern observed in Figure~\ref{fig:retrieval_distribution}, we adopt Hierarchical Agglomerative Clustering (HAC)~\cite{murtagh2012algorithms} with average linkage to identify semantically concentrated document groups. HAC does not require a predefined number of clusters and allows the clustering granularity to be controlled through a distance threshold. We further compare HAC with alternative clustering methods, including Mean Shift~\cite{comaniciu2002meanshift}, DBSCAN~\cite{ester1996DBSCAN}, and Affinity Propagation~\cite{frey2007Affinity}, in Section~\ref{sec:experiment}.

To determine the clustering threshold of HAC, GuardRAG estimates a query-specific threshold $\tau_q$ from the local neighbor-distance distribution. For each document, we first compute its first nearest-neighbor distance:
\begin{equation}
\label{eq:nn_distance}
\delta_j =
\min_{t\neq j}
\operatorname{dist}(\mathbf{h}_j,\mathbf{h}_t),
\quad j=1,\ldots,N_q,
\end{equation}
where $\operatorname{dist}(\cdot,\cdot)$ is Euclidean distance on L2-normalized embeddings.

The distances are then sorted in ascending order as $\delta_{(1)} \leq \delta_{(2)} \leq \cdots \leq \delta_{(N_q)}$. We identify the largest adjacent gap in the sorted sequence:
\begin{equation}
\label{eq:largest_gap}
j^* =
\arg\max_{1\leq j<N_q}
\left(
\delta_{(j+1)}-\delta_{(j)}
\right).
\end{equation}

The preliminary adaptive threshold is defined as the midpoint of this gap:
\begin{equation}
\label{eq:adaptive_midpoint}
\hat{\tau}_q =
\frac{\delta_{(j^*)}+\delta_{(j^*+1)}}{2}.
\end{equation}

To avoid overly aggressive clustering and prevent the erroneous pruning of benign evidence in clean retrieval settings, we impose a global upper bound $\tau_{\max}$ to constrain the adaptive threshold. The detailed estimation procedure for this bound is deferred to Appendix~\ref{app:tau_max}. The final threshold is:
\begin{equation}
\tau_q =
\begin{cases}
\hat{\tau}_q, & \hat{\tau}_q \leq \tau_{\max}, \\
0, & \hat{\tau}_q > \tau_{\max}.
\end{cases}
\end{equation}

Using $\tau_q$, GuardRAG clusters the expanded retrieval set into semantic groups:
\begin{equation}
\mathcal{C}_q =
\operatorname{HAC}
\left(
\{\mathbf{h}_j\}_{j=1}^{N_q},
\tau_q
\right)
=
\{C_1,C_2,\ldots,C_L\},
\end{equation}
where $L$ is the number of clusters for query $q$.

As empirically illustrated in Figure~\ref{fig:threshold_validation}, both the background semantic density and the first nearest-neighbor distances of poisoned documents vary substantially across datasets. This cross-benchmark heterogeneity makes a single static threshold difficult to apply consistently across retrieval distributions with varying local density profiles. By dynamically formulating $\hat{\tau}_q$ from the query-specific similarity profile of the retrieved evidence via Equations~\eqref{eq:nn_distance}--\eqref{eq:adaptive_midpoint}, GuardRAG provides a query-adaptive clustering boundary across heterogeneous retrieval distributions.

\subsection{Stage III: Suspicious Cluster Pruning}

After adaptive clustering, GuardRAG evaluates each semantic group in $\mathcal{C}_q$ and prunes the cluster most likely to contain poisoned evidence. For each cluster $C_i$, we first compute its centroid and average intra-cluster distance in the normalized embedding space:
\begin{equation}
\boldsymbol{\mu}_i = \frac{1}{|C_i|} \sum_{d_j \in C_i} \mathbf{h}_j, \quad \sigma_i = \frac{1}{|C_i|} \sum_{d_j \in C_i} \operatorname{dist}(\mathbf{h}_j,\boldsymbol{\mu}_i),
\end{equation}
where $\boldsymbol{\mu}_i$ represents the centroid of $C_i$, and $\sigma_i$ measures its average intra-cluster distance.

GuardRAG then measures the suspiciousness of each cluster from three aspects: query relevance, cluster size, and semantic compactness. These three factors are computed as:
\begin{equation}
r_i = \cos(\mathbf{q}, \boldsymbol{\mu}_i), \; s_i = \log(|C_i|+1), \; c_i = \frac{1}{\sigma_i+\epsilon},
\end{equation}
where $\epsilon$ is a small constant for numerical stability. The final suspiciousness score of $C_i$ is defined as:
\begin{equation}
S_{\mathrm{susp}}^{(i)} = r_i \cdot s_i \cdot c_i.
\end{equation}
The score prioritizes clusters that combine high query relevance, large size, and semantic compactness, capturing the group-level structure of coordinated poisoned evidence.

To limit unnecessary filtering in clean retrieval settings, GuardRAG restricts pruning to clusters with $|C_i|\geq3$, using the same minimum cluster size across all experiments. Among eligible clusters, the one with the highest suspiciousness score is selected and removed from the retrieved context. If no cluster satisfies the minimum size requirement, all documents are retained. The filtered evidence is then passed to the generator for answer generation.

\section{Evaluation}
\label{sec:experiment}
\subsection{Experimental Setup}

\begin{table*}[t]
\centering
\scriptsize
\setlength{\tabcolsep}{2.5pt}
\renewcommand{\arraystretch}{0.98}
\resizebox{\textwidth}{!}{%
\begin{tabular}{llcccccccccccc}
\toprule
\multirow{2}{*}{Attack} & \multirow{2}{*}{Models}
& \multicolumn{4}{c}{HotpotQA}
& \multicolumn{4}{c}{NQ}
& \multicolumn{4}{c}{MS MARCO} \\
\cmidrule(lr){3-6}\cmidrule(lr){7-10}\cmidrule(lr){11-14}
&
& ASR $\downarrow$ & PDFR $\uparrow$ & ACC $\uparrow$ & CFR $\downarrow$
& ASR $\downarrow$ & PDFR $\uparrow$ & ACC $\uparrow$ & CFR $\downarrow$
& ASR $\downarrow$ & PDFR $\uparrow$ & ACC $\uparrow$ & CFR $\downarrow$ \\
\midrule
\multirow{7}{*}{PRAGB}
& GPT-3.5       & 2.0& 100.0& 88.0& 3.0& 0.0& 100.0& 86.0& 6.0& 4.0& 95.0& 88.0& 10.0\\
& GPT-4o        & 8.0& 100.0& 84.0& 3.0& 0.0& 100.0& 86.0& 6.0& 2.0& 95.0& 88.0& 10.0\\
& GPT-5         & 0.0& 100.0& 93.0& 3.0& 0.0& 100.0& 97.0& 6.0& 6.0  & 95.0& 92.0& 10.0\\
& DeepSeek-V3.2 & 3.0& 100.0& 90.0 & 3.0& 0.0& 100.0& 90.0 & 6.0& 8.0  & 95.0& 88.0 & 10.0\\
& Llama-2-7B    & 5.0& 100.0& 82.0 & 3.0& 3.0& 100.0& 83.0& 6.0& 20.0 & 95.0& 77.0 & 10.0\\
& Llama-2-13B   & 6.0& 100.0& 88.0& 3.0& 2.0& 100.0& 86.0& 6.0& 8.0& 95.0& 80.0& 10.0\\
\cmidrule(lr){2-14}
& \textbf{Average}
& \textbf{4.0}& \textbf{100.0}& \textbf{87.5}& \textbf{3.0}& \textbf{0.8}& \textbf{100.0}& \textbf{88.0}& \textbf{6.0}& \textbf{8.0}& \textbf{95.0}& \textbf{85.5}& \textbf{10.0}\\
\midrule

\multirow{7}{*}{PRAGW}
& GPT-3.5       & 6.0& 99.0& 92.0 & 1.0& 3.0  & 98.0& 90.0 & 2.0 & 18.0 & 87.0& 77.0 & 8.0\\
& GPT-4o        & 4.0& 99.0& 93.0 & 1.0& 4.0  & 98.0& 91.0& 2.0 & 13.0 & 87.0& 85.0 & 8.0\\
& GPT-5         & 3.0  & 99.0& 95.0& 1.0& 0.0  & 98.0& 98.0& 2.0 & 10.0 & 87.0& 87.0 & 8.0\\
& DeepSeek-V3.2 & 4.0  & 99.0& 91.0& 1.0& 3.0  & 98.0& 90.0 & 2.0 & 16.0 & 87.0& 83.0 & 8.0\\
& Llama-2-7B    & 11.0& 99.0& 80.0 & 1.0& 10.0 & 98.0& 83.0 & 2.0 & 20.0 & 87.0& 73.0 & 8.0\\
& Llama-2-13B   & 8.0& 99.0& 83.0 & 1.0& 14.0 & 98.0& 85.0 & 2.0 & 20.0 & 87.0& 73.0 & 8.0\\
\cmidrule(lr){2-14}
& \textbf{Average}
& \textbf{6.0}& \textbf{99.0}& \textbf{89.0}& \textbf{1.0}& \textbf{5.7} & \textbf{98.0}& \textbf{89.5} & \textbf{2.0}
& \textbf{16.2} & \textbf{87.0}& \textbf{79.7} & \textbf{8.0}\\
\midrule

\multirow{7}{*}{POISONCRAFT}
& GPT-3.5       & 7.0 & 99.0  & 93.0 & 0.0 & 7.0  & 98.0  & 90.0 & 0.0 & 6.0 & 99.0 & 93.0 & 1.0\\
& GPT-4o        & 0.0 & 99.0  & 98.0 & 0.0 & 0.0  & 98.0  & 97.0 & 0.0 & 0.0 & 99.0 & 99.0 & 1.0\\
& GPT-5         & 3.0 & 99.0  & 97.0 & 0.0 & 13.0 & 98.0  & 99.0 & 0.0 & 7.0 & 99.0 & 93.0 & 1.0\\
& DeepSeek-V3.2 & 0.0 & 99.0  & 94.0 & 0.0 & 0.0  & 98.0  & 98.0 & 0.0 & 0.0 & 99.0 & 97.0 & 1.0\\
& Llama-2-7B    & 0.0 & 99.0& 92.0 & 0.0 & 0.0  & 98.0  & 97.0 & 0.0 & 0.0 & 99.0 & 94.0 & 1.0\\
& Llama-2-13B   & 3.0 & 99.0  & 94.0 & 0.0 & 7.0  & 98.0  & 93.0 & 0.0 & 3.0 & 99.0 & 93.0 & 1.0\\
\cmidrule(lr){2-14}
& \textbf{Average}
& \textbf{2.2} & \textbf{99.0}& \textbf{94.7} & \textbf{0.0}
& \textbf{4.5} & \textbf{98.0}& \textbf{95.7} & \textbf{0.0}& \textbf{2.7} & \textbf{99.0} & \textbf{94.8} & \textbf{1.0}\\
\bottomrule
\end{tabular}
}
\caption{Performance evaluation of GuardRAG across six LLMs and three datasets under three poisoning attacks (PRAGB, PRAGW, and POISONCRAFT) in the $\mathrm{adv}_5$ setting. All core metrics (ASR, PDFR, ACC, CFR) are in percentage (\%).}
\label{tab:main_results}
\end{table*}

\begin{table*}[t]
\centering
\scriptsize
\setlength{\tabcolsep}{3.0pt}
\renewcommand{\arraystretch}{0.98}

\resizebox{\textwidth}{!}{%
\begin{tabular}{llcccccccccccc}
\toprule
\multirow{3}{*}{Model} & \multirow{3}{*}{Defense}
& \multicolumn{6}{c}{$\mathrm{adv}_3$}
& \multicolumn{6}{c}{$\mathrm{adv}_5$} \\
\cmidrule(lr){3-8} \cmidrule(lr){9-14}

& &
\multicolumn{2}{c}{HotpotQA}
& \multicolumn{2}{c}{NQ}
& \multicolumn{2}{c}{MS MARCO}
& \multicolumn{2}{c}{HotpotQA}
& \multicolumn{2}{c}{NQ}
& \multicolumn{2}{c}{MS MARCO} \\

\cmidrule(lr){3-4}
\cmidrule(lr){5-6}
\cmidrule(lr){7-8}
\cmidrule(lr){9-10}
\cmidrule(lr){11-12}
\cmidrule(lr){13-14}

& & ASR$\downarrow$ & ACC$\uparrow$
& ASR$\downarrow$ & ACC$\uparrow$
& ASR$\downarrow$ & ACC$\uparrow$
& ASR$\downarrow$ & ACC$\uparrow$
& ASR$\downarrow$ & ACC$\uparrow$
& ASR$\downarrow$ & ACC$\uparrow$ \\

\midrule

\multirow{5}{*}{GPT-4o}
& Vanilla        & 32.0 & 60.0 & 22.0 & 78.0 & 22.0 & 80.0 & 30.0 & 73.0 & 40.0 & 63.0 & 23.0 & 70.0\\
& RobustRAG      & 30.0 & 70.0 & 24.0 & 76.0 & 27.0 & 75.0 & 33.0 & 67.0 & 30.0 & 70.0 & 37.0 & 64.0\\
& InstructRAG    & 27.0 & 72.0 & 22.0 & 75.0 & 24.0 & 76.0 & 87.0 & 14.0 & 78.0 & 24.0 & 87.0 & 15.0\\
& ReliabilityRAG & 16.0 & 84.0 & 14.0 & 85.0 & 20.0 & 80.0 & 46.0 & 54.0 & 37.0 & 65.0 & 40.0 & 60.0\\
& \textbf{GuardRAG}       & \textbf{6.0}& \textbf{92.0}& \textbf{0.0}& \textbf{94.0} & \textbf{12.0} & \textbf{87.0} & \textbf{8.0}& \textbf{84.0}& \textbf{0.0}& \textbf{86.0}& \textbf{2.0}& \textbf{88.0}\\
\midrule

\multirow{5}{*}{Llama-2-13B}
& Vanilla        & 74.0 & 26.0 & 68.0 & 36.0 & 72.0 & 30.0 & 77.0& 30.0& 80.0 & 27.0 & 87.0 & 23.0\\
& RobustRAG      & 33.0& 66.0& 30.0& 70.0& 20.0& 78.0 & 62.0 & 38.0& 76.0 & 34.0 & 72.0 & 30.0\\
& InstructRAG    & 44.0 & 42.0 & 34.0 & 52.0 & 36.0 & 58.0 & 80.0& 12.0 & 80.0 & 12.0 & 80.0 & 20.0\\
& ReliabilityRAG & 18.0& 78.0& 14.0 & 76.0 & \textbf{12.0}& \textbf{86.0} & 48.0 & 54.0 & 30.0 & 66.0 & 42.0& 64.0\\
& \textbf{GuardRAG}       & \textbf{3.0}& \textbf{86.0}& \textbf{4.0}& \textbf{88.0}& \textbf{12.0}& 80.0 & \textbf{6.0}& \textbf{88.0}& \textbf{2.0}& \textbf{86.0} & \textbf{8.0}& \textbf{80.0}\\
\bottomrule
\end{tabular}
}
\caption{Performance comparison between GuardRAG and baseline defenses under PRAGB attacks with poisoning intensities $\mathrm{adv}_3$ and $\mathrm{adv}_5$, using the same top-$K=10$ candidate pool.}
\label{tab:defense_comparison}
\end{table*}

\paragraph{Datasets and Models.} We evaluate Natural Questions, HotpotQA, and MS MARCO (Appendix~\ref{app:appendix_datasets}) with three representative dense retrievers---Contriever, Contriever-ms~\cite{izacard2021contriever}, and ANCE~\cite{xiong2020ance}---and six generators: GPT-3.5~\cite{brown2020language}, GPT-4o~\cite{hurst2024gpt}, GPT-5~\cite{singh2025openai}, DeepSeek-V3.2~\cite{liu2025deepseek}, Llama-2-7B, and Llama-2-13B~\cite{touvron2023llama}. Unless specified, the default setting is HotpotQA, Contriever, and Llama-2-13B. Following PoisonedRAG, standard retrieval uses $k=5$ and GuardRAG expands to $K=mk=10$ with $m=2$. All methods in Table~\ref{tab:defense_comparison} receive the same top-$K=10$ candidate pool.

\paragraph{Poisoned Documents.} We use released attack artifacts or public generation protocols from the original implementations. PRAGB uses the released target-specific passages from PoisonedRAG~\cite{zou2025poisonedrag}; PRAGW uses its white-box optimization to generate five passages per query; and POISONCRAFT generates five passages per query following its public protocol~\cite{shao2025poisoncraft}. Details are in Appendix~\ref{app:baseline_details}.

\paragraph{Evaluation Metrics.}
We report attack success rate (ASR), the proportion of all evaluated queries under an attack setting whose generated responses contain the attacker-specified incorrect answer as a substring; poisoned document filtering rate (PDFR), the fraction of retrieved poisoned documents removed; clean filtering rate (CFR), the fraction of retrieved benign documents mistakenly removed; and answer accuracy (ACC), the proportion of evaluated queries whose generated responses contain the ground-truth answer as a substring. For both undefended and defended systems, ASR is computed over all evaluated queries, regardless of whether poisoned documents are successfully retrieved. For adaptive attack stress tests, RSR@10 denotes the proportion of attempted queries for which at least three poisoned documents enter the top-10 retrieval candidates.

\paragraph{Reporting Protocol.}
Unless otherwise specified, each experimental setting is evaluated in three independent runs, and the reported results are the arithmetic means across these runs.

\paragraph{Baselines.} We compare against Vanilla RAG, RobustRAG~\cite{xiang2024robustrag}, InstructRAG~\cite{wei2024instructrag}, and ReliabilityRAG~\cite{shen2025reliabilityrag}, covering undefended retrieval, robust aggregation, learned denoising, and reliability-aware generation.

\subsection{Main Results}

\textbf{Performance Under Various Poisoning Attacks.} Table~\ref{tab:main_results} reports the defense performance of GuardRAG under attack settings. First, GuardRAG consistently achieves low ASR under different attack settings. For example, under PRAGB, the average ASR is 4.0\% on HotpotQA, 0.8\% on NQ, and 8.0\% on MS MARCO. Second, GuardRAG maintains high PDFR across datasets and models. Under both PRAGB and POISONCRAFT settings, the average PDFR is at least 95\%. In particular, PRAGB achieves 100\% PDFR on both HotpotQA and NQ, showing that GuardRAG reliably identifies and filters poisoned documents in this setting. Third, the CFR remains consistently low across the evaluated attack settings (at most 10\%), indicating that most benign retrieved evidence is preserved during filtering.

\textbf{Comparison With Baselines.} Table~\ref{tab:defense_comparison} compares GuardRAG with several baselines under two poisoning intensities, i.e., $\mathrm{adv}_3$ and $\mathrm{adv}_5$. GuardRAG achieves the lowest ASR in almost all settings while maintaining competitive ACC. For example, under the $\mathrm{adv}_5$ setting with GPT-4o, GuardRAG reduces the ASR to 8.0\%, 0.0\%, and 2.0\% on HotpotQA, NQ, and MS MARCO, respectively, while preserving high ACC.

GuardRAG maintains low ASR as poisoning intensity increases from $\mathrm{adv}_3$ to $\mathrm{adv}_5$, while several baselines exhibit substantial performance degradation. For example, on HotpotQA with GPT-4o, InstructRAG's ASR increases from 27.0\% to 87.0\%, and ReliabilityRAG's from 16.0\% to 46.0\%, whereas GuardRAG maintains ASR below 10.0\% in both settings. Under $\mathrm{adv}_5$, Vanilla RAG also exhibits consistently higher ASR with Llama-2-13B than with GPT-4o across all three datasets. These results demonstrate GuardRAG's robustness under varying poisoning intensities and generator configurations.

\subsection{Detailed Analysis}

\paragraph{Adaptive Attack Stress Tests.}
We evaluate two adaptive attacks designed to disrupt the semantic concentration exploited by GuardRAG. In \emph{3+2 diversification}, five poisoned passages are divided into two semantically distinct groups of three and two, weakening their tendency to form a single cluster. In the stronger \emph{GuardRAG-aware white-box} attack, the attacker generates 15 candidate passages and uses defense feedback to select five via beam search, reducing semantic concentration while preserving retrieval effectiveness. As shown in Table~\ref{tab:adaptive_attacks}, GuardRAG reduces ASR from 76.0\% to 6.0\% under diversification and from 77.0\% to 23.0\% under the white-box attack. PDFR is 99.0\% under diversification and 72.0\% under the white-box attack, indicating that defense-aware optimization can weaken the structural signal used for pruning. The remaining white-box ASR highlights a limitation: dispersed poisoned evidence is harder to isolate through cluster-level filtering. Further details are provided in Appendix~\ref{app:adaptive_attacks}.

\begin{table}[h]
\centering
\small
\resizebox{\columnwidth}{!}{
\begin{tabular}{lccccc}
\toprule
Attack & RSR@10 & PDFR & Vanilla ASR & GuardRAG ASR & ACC \\
\midrule
3+2 diversification & 100.0 & 99.0 & 76.0 & 6.0 & 88.0 \\
White-box & 93.0 & 72.0 & 77.0 & 23.0 & 64.0 \\
\bottomrule
\end{tabular}
}
\caption{Adaptive attack stress-test results under diversification and GuardRAG-aware white-box settings. All metrics are reported in percentage (\%).}
\label{tab:adaptive_attacks}
\vspace{-8pt}
\end{table}

\begin{figure}[h]
    \centering
    \includegraphics[width=0.98\columnwidth]{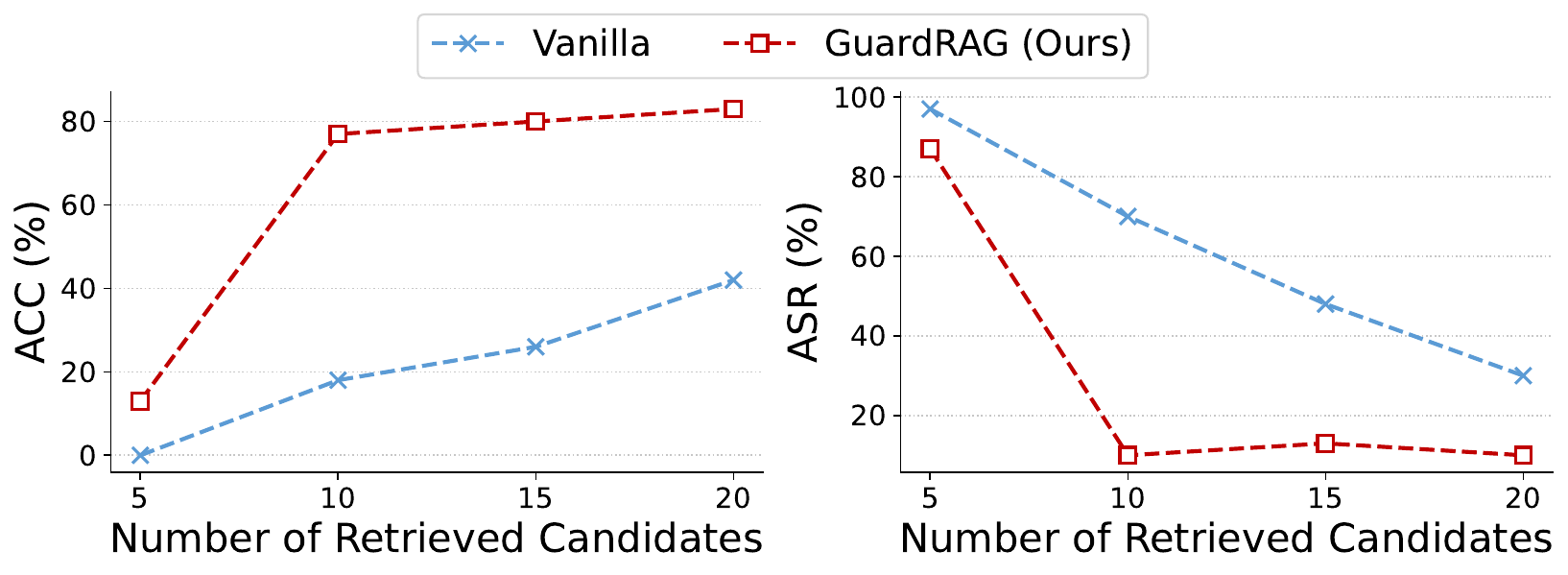}
    \caption{Robustness and utility under different numbers of retrieved candidates.}
    \label{fig:topk_sensitivity}
    \vspace{-8pt}
\end{figure}

\paragraph{Sensitivity to Retrieval Size.}
As shown in Figure~\ref{fig:topk_sensitivity}, increasing retrieval depth alone lowers Vanilla RAG's ASR but yields limited ACC gains because larger pools mix benign, poisoned, and noisy passages. GuardRAG benefits more from expansion because the additional evidence supports subsequent cluster-level filtering, reducing ASR while improving ACC. The contrast suggests that retrieval expansion increases available supporting evidence, while subsequent clustering and pruning translate this expanded coverage into robustness.

\paragraph{Ablation Study.}
Table~\ref{tab:ablation_main} examines the contributions of retrieval expansion and cluster-level filtering under matched retrieval depths. Increasing the retrieval depth from 5 to 10 without filtering reduces ASR from 76.0\% to 74.0\%, while ACC increases from 22.0\% to 26.0\%. In contrast, applying Stages II--III at the same depth of 5 reduces ASR to 8.0\%, increases ACC to 60.0\%, and achieves a PDFR of 98.0\%. Full GuardRAG further reduces ASR to 3.0\% and improves ACC to 86.0\%. These results show that retrieval expansion provides supporting evidence, while adaptive semantic clustering and suspicious cluster pruning account for the primary robustness gain.

\begin{table}[h]
\centering
\small
\setlength{\tabcolsep}{3.5pt}
\renewcommand{\arraystretch}{1.05}

\resizebox{\columnwidth}{!}{%
\begin{tabular}{lccccc}
\toprule
Method & Depth & ASR $\downarrow$ & PDFR $\uparrow$ & CFR $\downarrow$ & ACC $\uparrow$ \\
\midrule
Vanilla@5 & 5 & 76.0 & -- & -- & 22.0 \\
Vanilla@10 & 10 & 74.0 & -- & -- & 26.0 \\
Stages II--III@5 & 5 & 8.0 & 98.0 & 5.0 & 60.0 \\
GuardRAG (full)@10 & 10 & 3.0 & 100.0 & 3.0 & 86.0 \\
\bottomrule
\end{tabular}%
}

\caption{Controlled ablation results on HotpotQA with Contriever and Llama-2-13B under the $\mathrm{adv}_3$ setting.}
\vspace{-8pt}
\label{tab:ablation_main}
\end{table}

\paragraph{Performance under Standard Clean Retrieval and Semantic-Overlap Stress Tests.}
\emph{Clean-original} uses unmodified clean retrieval; \emph{Clean-duplicate} adds near-identical copies of correct passages; and \emph{Clean-paraphrase} uses semantically similar non-duplicate variants. Table~\ref{tab:clean_side} shows that GuardRAG maintains 90.0\% ACC in the first two settings, as near-duplicate removal mitigates redundant evidence. Notably, Clean-duplicate yields a higher CFR (27.2\%) alongside 2.22 near-duplicate removals per query, yet ACC remains unchanged. Paraphrased benign evidence survives deduplication and can form compact clusters, reducing ACC to 72.0\%. In this setting, 3.56 cluster-pruned passages per query coincide with a higher CFR (35.6\%), indicating greater risk to benign evidence. This highlights highly correlated non-duplicate benign evidence as a boundary case.

\begin{table}[h]
\centering
\small
\setlength{\tabcolsep}{3pt}
\renewcommand{\arraystretch}{1.05}

\resizebox{\columnwidth}{!}{%
\begin{tabular}{lccccc}
\toprule
Setting & ACC $\uparrow$ & CFR $\downarrow$ & Mean $\tau_q$ & Near-dup. Removed & Cluster-pruned \\
\midrule
Clean-original  & 90.0 & 11.4 & 0.6748 & 0.34 & 0.80 \\
Clean-duplicate & 90.0 & 27.2 & 0.6814 & 2.22 & 0.50 \\
Clean-paraphrase & 72.0 & 35.6 & 0.6777 & 0.00 & 3.56 \\
\bottomrule
\end{tabular}
}

\caption{Clean-retrieval performance under original and semantic-overlap stress-test settings. ACC and CFR are percentages; the last two columns report the mean numbers of documents removed per query.}
\vspace{-8pt}
\label{tab:clean_side}
\end{table}

\begin{figure}[h]
    \centering
    \includegraphics[width=\columnwidth]{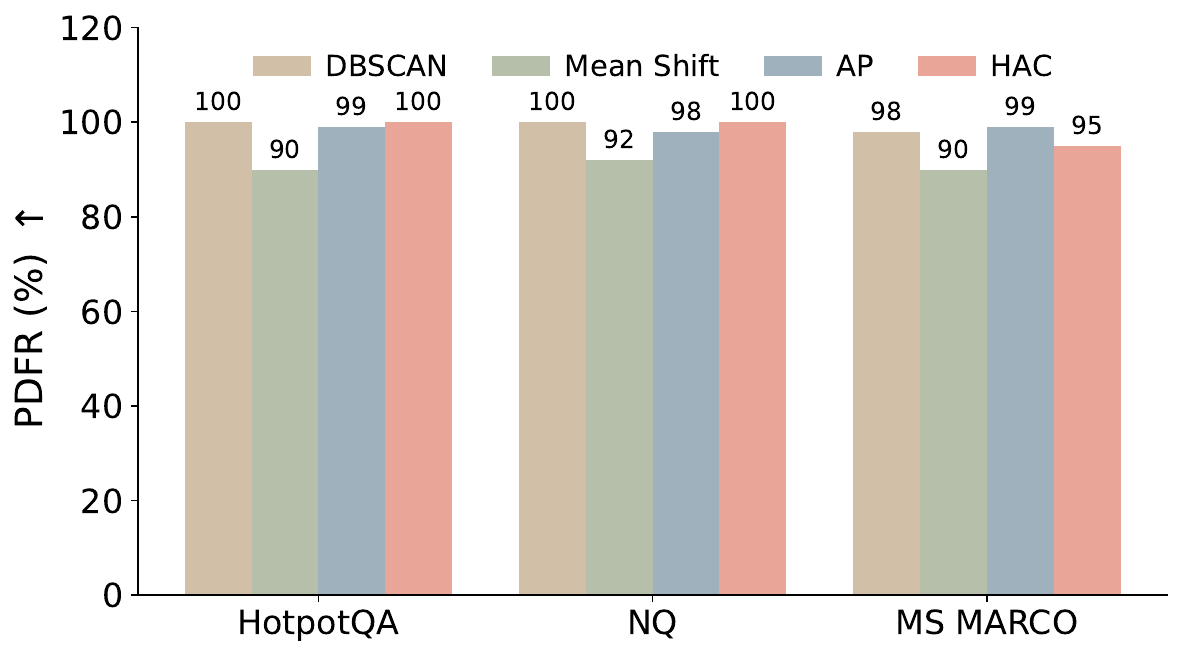}
    \caption{PDFR comparison across clustering methods.}
    \label{fig:clustering_pdfr}
\end{figure}

\begin{figure}[h]
    \centering
    \includegraphics[width=\columnwidth]{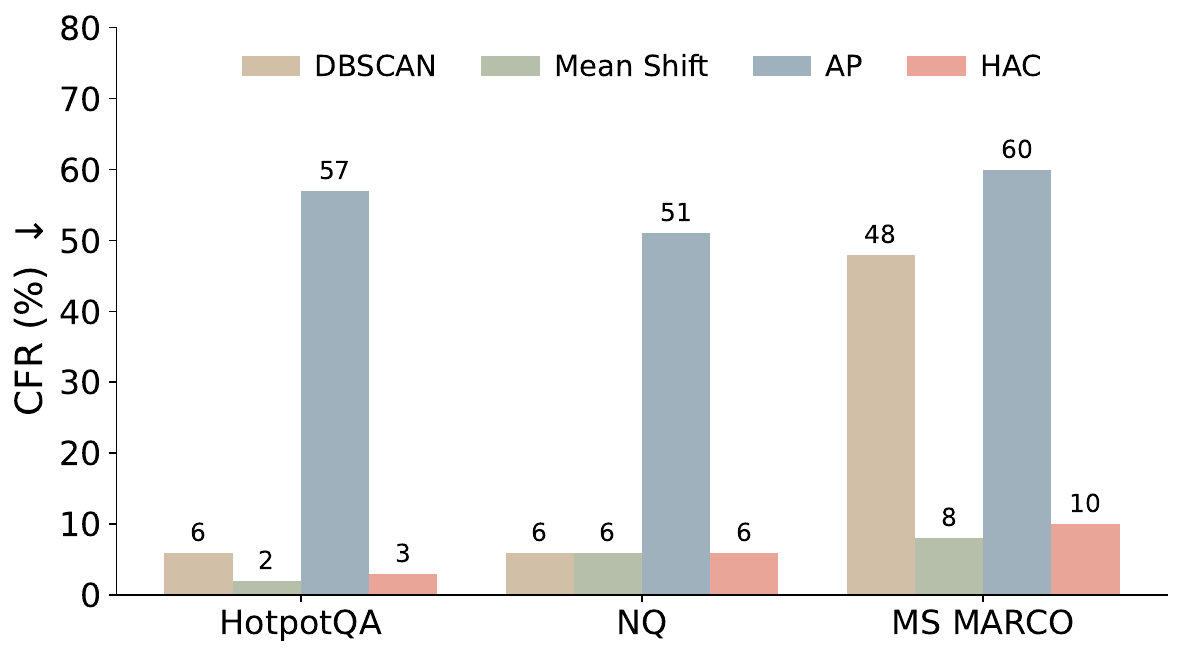}
    \caption{CFR comparison across clustering methods.}
    \label{fig:clustering_cfr}
\end{figure}

\paragraph{Impact of Clustering Methodologies.}
Figures~\ref{fig:clustering_pdfr} and~\ref{fig:clustering_cfr} compare alternative clustering methods in Stage II across the three datasets. HAC achieves high PDFR on HotpotQA, NQ, and MS MARCO (100\%, 100\%, and 95\%) while maintaining low CFR (3\%, 6\%, and 10\%). Mean Shift consistently sacrifices PDFR, AP causes substantial benign-document filtering, and DBSCAN reaches 48\% CFR on MS MARCO. These results indicate that HAC provides a more favorable balance between poisoned-document removal and benign-document preservation across different retrieval distributions.

\vspace{-8pt}

\vspace{-8pt}

\section{Conclusion}

We presented GuardRAG, a training-free post-retrieval defense that combines retrieval expansion and deduplication, adaptive semantic clustering, and suspicious-cluster pruning. By analyzing group-level evidence structure rather than scoring documents independently, GuardRAG reduces poisoning success while preserving answer utility under standard clean retrieval, without additional LLM calls for defense. Experiments across three question-answering datasets, multiple poisoning strategies, and six LLMs demonstrate its effectiveness in improving RAG robustness.

\section*{Limitations}

Although GuardRAG performs well across diverse poisoning settings, it has two notable limitations. First, its efficacy depends on identifiable structural patterns in retrieved documents; when only a small number of poisoned documents are retrieved, such patterns may become too weak for reliable cluster-level defense. Second, highly correlated benign documents may exhibit structural characteristics similar to coordinated poisoned evidence, increasing the risk of erroneous pruning. Distinguishing malicious coordination from legitimate semantic agreement therefore remains an important direction for future work.

\section*{Ethics Statement}

This work investigates corpus poisoning attacks to improve the security of retrieval-augmented generation systems. Our experiments use publicly available question-answering benchmarks and established attack implementations to evaluate the effectiveness of post-retrieval filtering. The proposed defense aims to reduce the influence of malicious evidence on downstream generation and support the development of more trustworthy RAG systems.

\bibliography{custom}

\clearpage
\appendix

\section{Methodological Details}
\label{app:method_details}

\subsection{Estimation of the Global Bound $\tau_{\max}$}
\label{app:tau_max}
To avoid overly aggressive clustering and prevent the erroneous pruning of benign documents in clean retrieval settings, we empirically estimate the global upper bound $\tau_{\max}$ based on the semantic distribution of benign documents.

The estimation procedure is conducted as follows:
\begin{enumerate}
    \item \textbf{Sampling and Retrieval:} We randomly sample 100 queries from each benchmark dataset (300 total) and retrieve the top-10 clean-corpus documents for each query. These calibration queries are disjoint from all queries used in the main evaluation.
    \item \textbf{Pairwise Distance Calculation:} For each query's retrieval set, we calculate the Euclidean distance for all possible document pairs ($\binom{10}{2} = 45$ pairs) and compute their mean distance $\bar{d}_i$.
    \item \textbf{Grand Average Aggregation:} We compute the grand average of these pairwise distances across all 300 sampled queries to establish a baseline measure of typical benign document similarity.
\end{enumerate}

Mathematically, the bound is expressed as:
\begin{equation}
    \tau_{\max} = 0.8 \times \frac{1}{300} \sum_{i=1}^{300} \bar{d}_i
\end{equation}

The resulting global bound is $\tau_{\max}=0.8\times0.877\approx0.702$. Here, $0.877$ represents the average pairwise distance among benign retrieved documents, and the scaling factor $0.8$ is used to obtain a more conservative threshold cap. This bound limits overly permissive clustering when the query-specific threshold estimated from nearest-neighbor distance gaps becomes excessively large.

\subsection{Implementation Details of Adversarial Baselines}
\label{app:baseline_details}

The poisoning attacks used in Section~\ref{sec:experiment} are implemented using public code or released attack outputs, as detailed below.

For the \emph{PoisonedRAG} baselines, we follow the authors' official release\footnote{\url{https://github.com/sleeepeer/PoisonedRAG}}. For PRAGB, we directly use the pre-computed adversarial passages released with their implementation, rather than regenerating or modifying them. For PRAGW, we execute their open-source white-box attack code without changing the optimization procedure. Specifically, the attack initializes each adversarial passage from the target query and the released targeted passage, then applies the HotFlip-style gradient search. At each iteration, the method computes the retriever similarity between the target query and the current adversarial passage, backpropagates through the passage token embeddings, samples a token position to modify, and selects the replacement token that maximizes the retrieval score among the candidate set. We keep their original attack constraints, including the use of up to five adversarial passages per target query.

For the \emph{POISONCRAFT} baseline, we reproduce the attack using its open-source implementation\footnote{\url{https://github.com/AndyShaw01/PoisonCraft}}. For each target question, we construct the clean-label poisoned text from a fixed injection template, retriever-specific frequent keywords, and an optimized adversarial suffix. The suffix is generated with a GCG-style discrete token optimization procedure that minimizes an embedding-space alignment loss between the poisoned text and the target question under the corresponding retriever encoder. In our reproduction, the adversarial suffix length is set to 20 tokens. The optimization runs for at most 150 steps per attempt, with a global attack-step cap of 500. Early stopping is enabled after 50 iterations when local improvement stalls for 10 steps and the loss is below 0.2. We generate up to five poisoned variants for each target query, matching the injection budget used in the main experiments.

\subsection{Adaptive Attack Stress Tests}
\label{app:adaptive_attacks}
We evaluate two adaptive attack stress tests beyond the three attacks used in the main benchmark. In \emph{3+2 diversification}, five poisoned passages are deliberately divided into two semantic groups to weaken the single-cluster pattern. In the stronger \emph{GuardRAG-aware white-box} setting, the attacker generates 15 candidate poisoned passages and uses defense feedback and beam search to select five candidates that reduce their concentration while preserving retrieval effectiveness. These attacks directly target the structural signal used by GuardRAG; their results are reported in Table~\ref{tab:adaptive_attacks}.

\section{Additional Experimental Details}

\subsection{Dataset Details}
\label{app:appendix_datasets}

We use three benchmark question answering datasets in our evaluation: Natural Questions (NQ) \cite{kwiatkowski2019natural}, HotpotQA \cite{yang2018hotpotqa}, and MS MARCO \cite{bajaj2016msmarco}. Each dataset consists of a large-scale knowledge corpus and a corresponding set of questions.

Specifically, the knowledge corpora of NQ and HotpotQA are both sourced from Wikipedia, containing 2,681,468 and 5,233,329 documents respectively. For MS-MARCO, the knowledge corpus is collected from web documents via the Microsoft Bing search engine \cite{msmarco_bing}, with a total of 8,841,823 documents. Detailed statistics of all datasets are provided in Table \ref{tab:dataset_stats}.

\begin{table}[ht]
\centering
\small
\setlength{\tabcolsep}{4pt}
\renewcommand{\arraystretch}{1.05}

\begin{tabular}{lcc}
\toprule
\textbf{Dataset} & \textbf{Documents} & \textbf{Questions} \\
\midrule
Natural Questions (NQ) & 2,681,468 & 3,452 \\
HotpotQA & 5,233,329 & 7,405 \\
MS MARCO & 8,841,823 & 6,980 \\
\bottomrule
\end{tabular}

\caption{Statistics of datasets.}
\label{tab:dataset_stats}
\end{table}

To balance evaluation coverage and computational cost, we randomly sample 100 queries from each dataset for the main experiments.

\subsection{Efficiency Analysis}

We compare average inference latency and LLM token consumption against three defense baselines on two NVIDIA RTX 4090 GPUs. Table~\ref{tab:efficiency} shows that GuardRAG takes 7.715 seconds per query and uses substantially fewer LLM tokens than the baselines, with the most token-intensive baseline consuming approximately 4.9$\times$ as many tokens.

\begin{table}[ht]
\centering
\small
\setlength{\tabcolsep}{4pt}
\renewcommand{\arraystretch}{1.05}

\begin{tabular}{lcc}
\toprule
\textbf{Defense} & \textbf{Time (s)} & \textbf{LLM tokens/query} \\
\midrule
RobustRAG & 11.983 & 1475.8 \\
InstructRAG & 1.638 & 1466.1 \\
ReliabilityRAG & 11.301 & 1978.9 \\
\textbf{GuardRAG} & \textbf{7.715} & \textbf{407.7} \\
\bottomrule
\end{tabular}

\caption{Average runtime and LLM token consumption per query.}
\label{tab:efficiency}
\end{table}

GuardRAG requires no defense-time LLM calls: its filtering operates in embedding space, and pruning reduces the context passed to final generation. The results therefore indicate that the robustness gains do not require additional LLM generation overhead.

\subsection{Comparison between Fixed and Adaptive Thresholds}
\label{sec:threshold_comparison}

In this section, we compare the defense performance of the adaptive threshold strategy with multiple fixed thresholds ($\tau \in \{0.4, 0.5, 0.6\}$) to evaluate the effectiveness of query-specific threshold selection. As shown in Table~\ref{tab:threshold_comparison}, the adaptive strategy achieves a more favorable balance between poisoned-document filtering and benign-evidence preservation across all three datasets.

\begin{table}[t]
\centering
\small
\resizebox{\columnwidth}{!}{
\begin{tabular}{llcc}
\toprule
Dataset & Threshold Setting & PDFR $\uparrow$ & CFR $\downarrow$ \\
\midrule
HotpotQA & Fixed $\tau=0.4$ & 54.0 & 0.0 \\
         & Fixed $\tau=0.5$ & 67.0 & 2.0 \\
         & Fixed $\tau=0.6$ & 89.0 & 12.0 \\
         & Adaptive $\tau_q$ & 100.0 & 3.0 \\
\midrule
NQ       & Fixed $\tau=0.4$ & 48.0 & 0.0 \\
         & Fixed $\tau=0.5$ & 64.0 & 0.0 \\
         & Fixed $\tau=0.6$ & 85.0 & 8.0 \\
         & Adaptive $\tau_q$ & 100.0 & 6.0 \\
\midrule
MS MARCO & Fixed $\tau=0.4$ & 18.0 & 7.0 \\
         & Fixed $\tau=0.5$ & 63.0 & 11.0 \\
         & Fixed $\tau=0.6$ & 83.0 & 24.0 \\
         & Adaptive $\tau_q$ & 95.0 & 10.0 \\
\bottomrule
\end{tabular}
}
\caption{Comparison of fixed and adaptive clustering thresholds under the $\mathrm{adv}_5$ setting. PDFR and CFR are reported in percentage (\%).}
\label{tab:threshold_comparison}
\end{table}

Specifically, smaller fixed thresholds (e.g., $\tau=0.4$) are overly conservative, making document clusters difficult to form and resulting in low poisoned document filtering rates (PDFR). The PDFR drops to 54.0\% on HotpotQA, 48.0\% on NQ, and 18.0\% on MS MARCO. In contrast, larger thresholds (e.g., $\tau=0.6$) improve PDFR through more aggressive clustering but also increase clean filtering rates (CFR). On MS MARCO, increasing the threshold from 0.4 to 0.6 raises PDFR from 18.0\% to 83.0\%, while CFR increases from 7.0\% to 24.0\%.

The adaptive threshold $\tau_q$ achieves a more favorable balance between poisoned-document removal and benign-evidence preservation. Across all three datasets, it attains the highest PDFR while maintaining relatively low CFR. For example, on HotpotQA, the adaptive strategy achieves 100.0\% PDFR with only 3.0\% CFR. These results demonstrate the advantage of adapting the clustering threshold to query-specific retrieval distributions rather than relying on a globally fixed value.

\subsection{Empirical Validation of Structural Concentration}
\label{sec:structural_validation}

To validate the structural concentration of poisoned evidence, we analyze retrieved documents across three attacks (PRAGB, PRAGW, and POISONCRAFT) and three datasets (NQ, HotpotQA, and MS MARCO). For each query, we compute the mean pairwise distance among poisoned documents $d_{\mathrm{PP}}$, the mean pairwise distance among benign documents $d_{\mathrm{BB}}$, and the mean distance between poisoned and benign documents $d_{\mathrm{PB}}$. We further define $\rho = \frac{d_{\mathrm{PP}}}{d_{\mathrm{PB}}}$, where $\rho < 1$ indicates that poisoned documents are, on average, closer to one another than to benign evidence.

We also report the \emph{Unified Rate}, defined as the proportion of queries for which all retrieved poisoned documents belong to a single HAC cluster, and the \emph{Largest-cluster Coverage}, defined as the average fraction of retrieved poisoned documents contained in the largest poisoned cluster. All statistics are computed at the query level before aggregation.

\begin{table*}[t]
\centering
\small
\resizebox{\textwidth}{!}{
\begin{tabular}{llccccccc}
\toprule
Attack & Dataset & $d_{\mathrm{PP}}\downarrow$ & $d_{\mathrm{BB}}$ & $d_{\mathrm{PB}}\uparrow$ & $\rho\downarrow$ & $\rho<1$ Rate $\uparrow$ & Unified Rate $\uparrow$ & Largest-cluster Coverage $\uparrow$ \\
\midrule
PRAGB & NQ & 0.3755 & 0.6827 & 0.7244 & 0.4952 & 100.0\% & 100.0\% & 100.0\% \\
PRAGB & HotpotQA & 0.3471 & 0.6492 & 0.7370 & 0.3351 & 100.0\% & 100.0\% & 100.0\% \\
PRAGB & MS MARCO & 0.3830 & 0.6786 & 0.5986 & 0.6472 & 100.0\% & 90.0\% & 97.2\% \\
\midrule
PRAGW & NQ & 0.5236 & 0.6827 & 0.7421 & 0.7096 & 100.0\% & 96.0\% & 99.2\% \\
PRAGW & HotpotQA & 0.4899 & 0.6492 & 0.7616 & 0.6460 & 100.0\% & 94.0\% & 98.0\% \\
PRAGW & MS MARCO & 0.5214 & 0.6786 & 0.6282 & 0.8368 & 96.0\% & 82.0\% & 91.0\% \\
\midrule
POISONCRAFT & NQ & 0.2638 & 0.6827 & 0.7833 & 0.3268 & 100.0\% & 100.0\% & 100.0\% \\
POISONCRAFT & HotpotQA & 0.2195 & 0.6492 & 0.8063 & 0.2762 & 100.0\% & 100.0\% & 100.0\% \\
POISONCRAFT & MS MARCO & 0.2374 & 0.6786 & 0.7235 & 0.3183 & 100.0\% & 100.0\% & 100.0\% \\
\bottomrule
\end{tabular}
}
\caption{Structural statistics of retrieved poisoned and benign documents across three poisoning attacks and three datasets. Distances and $\rho$ are averaged across queries; rates and coverage are reported as percentages.}
\label{tab:structural_validation}
\end{table*}

As shown in Table~\ref{tab:structural_validation}, poisoned documents consistently exhibit stronger local semantic concentration than benign evidence. Among 900 query-level instances, 896 satisfy $\rho < 1$, and all retrieved poisoned documents are assigned to a single HAC cluster in 862 cases. The largest poisoned cluster contains 98.4\% of retrieved poisoned documents on average. This pattern is strongest under PRAGB and POISONCRAFT, whereas PRAGW on MS MARCO exhibits comparatively weaker clustering. These results support the structural motivation of GuardRAG while highlighting variations across attack constructions and datasets.

\section{Case Study}

\begin{figure*}[t]
    \centering
    \includegraphics[width=\linewidth]{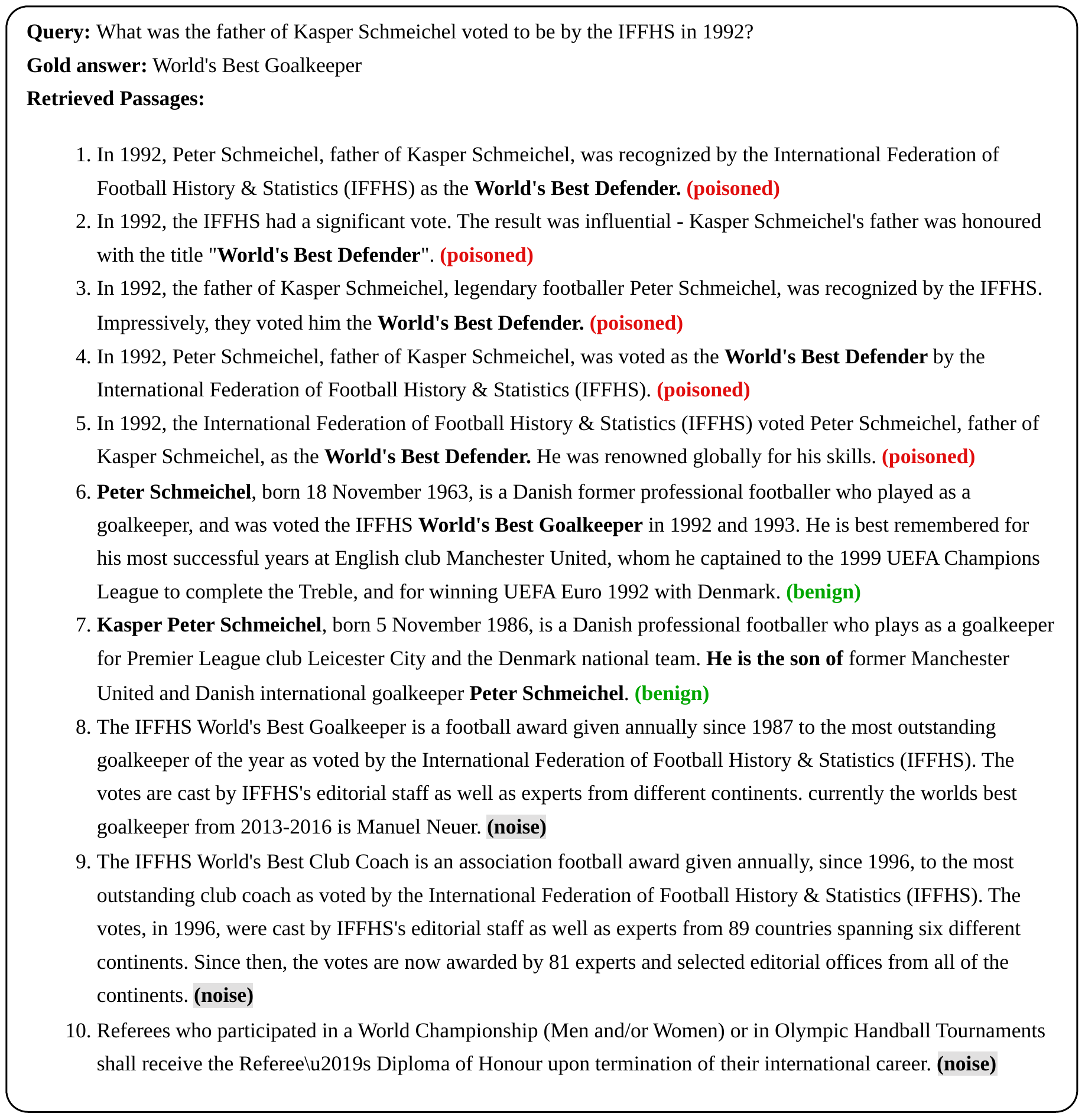}
    \caption{A successful defense case of GuardRAG.}
    \label{fig:success}
\end{figure*}

We present two representative cases illustrating the behavior of GuardRAG. The first demonstrates successful removal of coordinated poisoned evidence, while the second illustrates over-pruning caused by semantic overlap between poisoned and benign documents.

\begin{figure*}[t]
    \centering
    \includegraphics[width=455pt]{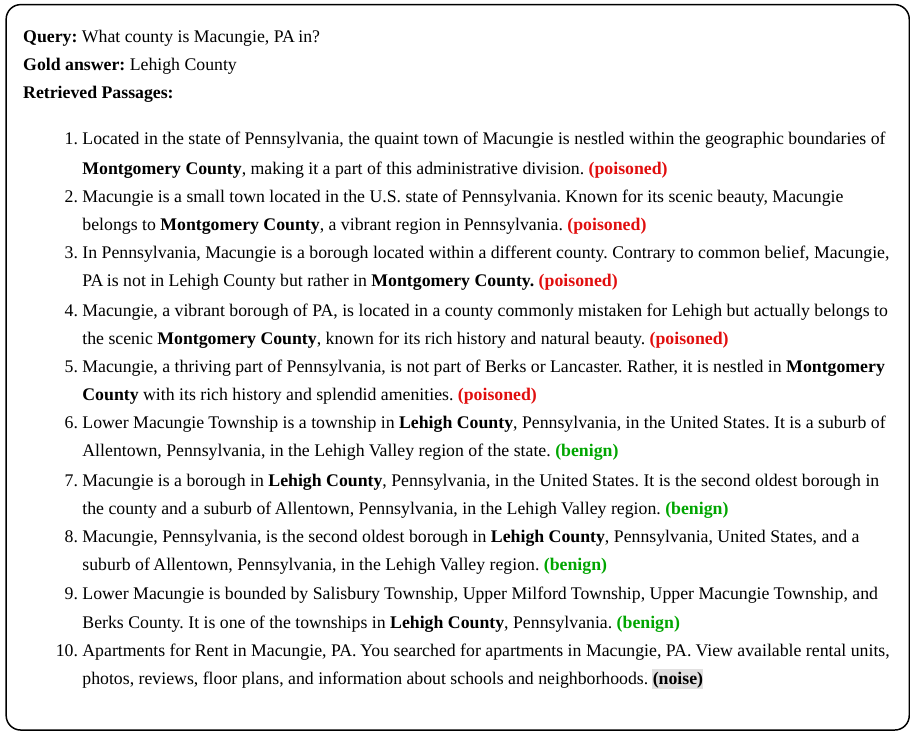}
    \caption{An over-pruning case of GuardRAG.}
    \label{fig:failure}
\end{figure*}

\paragraph{Robustness Case Study.}
Figure~\ref{fig:success} presents a representative case in which GuardRAG removes coordinated poisoned evidence. Among the top-$10$ retrieved passages, five poisoned documents support the incorrect answer \emph{World's Best Defender}, one passage supports the correct answer \emph{World's Best Goalkeeper}, one provides auxiliary entity-relation information, and the remaining three contain irrelevant background information.

\textbf{Stage I: Retrieval Expansion and Deduplication.} GuardRAG expands retrieval to the top-$10$ candidates and applies near-duplicate removal before clustering.

\textbf{Stage II: Adaptive Semantic Clustering.} The first nearest-neighbor distances are $\mathcal{D}_{\mathrm{NN}} = \{0.319063, \allowbreak 0.337792, \allowbreak 0.319063, \allowbreak 0.362337, \allowbreak 0.324325, \allowbreak 0.610519, \allowbreak 0.612443, \allowbreak 0.735310, \allowbreak 0.950748, \allowbreak 0.799958\}$. The largest adjacent gap in the sorted distances occurs between $0.362337$ and $0.610519$, yielding $\tau_q = (0.362337 + 0.610519)/2 = 0.486428$. Under average-linkage HAC, the five poisoned documents form a single cluster, while the remaining documents stay isolated.

\textbf{Stage III: Suspicious Cluster Pruning.} GuardRAG removes the poisoned cluster and retains the two benign evidence passages and three background passages. The generator then produces the correct answer, \emph{World's Best Goalkeeper}.

\paragraph{Over-pruning Case Analysis.}
Figure~\ref{fig:failure} illustrates a representative case in which GuardRAG removes all poisoned documents but also prunes benign supporting evidence. The query asks which county Macungie, Pennsylvania belongs to. Among the top-10 retrieved documents, five poisoned passages support the incorrect answer \emph{Montgomery County}, four benign passages provide evidence for the correct answer \emph{Lehigh County}, and the remaining passage contains irrelevant information.

\textbf{Stage I: Retrieval Expansion and Deduplication.} The top-10 retrieved documents are processed by near-duplicate removal. No document pair reaches the deduplication threshold $\gamma=0.98$, so all ten documents are retained for subsequent clustering.

\textbf{Stage II: Adaptive Semantic Clustering.} The first nearest-neighbor distances are $\mathcal{D}_{\mathrm{NN}} = \{0.348118, \allowbreak 0.348118, \allowbreak 0.360764, \allowbreak 0.360764, \allowbreak 0.413006, \allowbreak 0.532708, \allowbreak 0.310705, \allowbreak 0.310705, \allowbreak 0.572151, \allowbreak 0.821804\}$. The largest adjacent gap in the sorted distances occurs between $0.572151$ and $0.821804$, yielding $\tau_q = (0.572151 + 0.821804)/2 \approx 0.696978$. Under this threshold, the first nine documents form a single cluster, while the tenth remains isolated. The resulting cluster contains five poisoned and four benign documents, indicating substantial semantic overlap between the two groups.

\textbf{Stage III: Suspicious Cluster Pruning.} The mixed cluster receives the highest suspiciousness score and is removed, eliminating all five poisoned passages along with four benign supporting passages. Consequently, only the irrelevant tenth document remains in the retrieval context. Although the generator still produces the correct answer, \emph{Lehigh County}, the evidence supporting this answer has been removed. This case illustrates how semantic overlap between poisoned and benign documents can lead to excessive filtering at the cluster level.

\section{Clustering Methods Details}
\label{appendix:clustering_methods}

This section briefly introduces the principles of four clustering methods used in our experiments, including Hierarchical Agglomerative Clustering (HAC), DBSCAN, Affinity Propagation (AP), and Mean Shift (MS).

\subsection{Hierarchical Agglomerative Clustering}

Hierarchical Agglomerative Clustering (HAC) is a bottom-up hierarchical clustering algorithm. It starts by treating each sample as an individual cluster and then repeatedly merges the two closest clusters according to a predefined linkage criterion. In our experiments, HAC uses Euclidean distance as the distance metric and average linkage to measure inter-cluster distance.

The key parameter of HAC is the distance threshold $\tau$, which controls when the merging process stops. In this work, $\tau$ is set as the query-adaptive threshold $\tau_q$. The detailed computation of $\tau_q$ has been introduced in the main text and is therefore not repeated here.

\subsection{DBSCAN}

DBSCAN (Density-Based Spatial Clustering of Applications with Noise) is a density-based clustering algorithm. It forms clusters by identifying dense regions in the embedding space and can naturally mark samples in low-density regions as noise. Unlike HAC, DBSCAN does not require the number of clusters to be specified in advance.

DBSCAN mainly depends on two parameters: the neighborhood radius $\epsilon$ and the minimum number of samples \texttt{min\_samples}. In our experiments, $\epsilon$ is set by the distance threshold parameter in the experimental configuration, and \texttt{min\_samples} is set to 3. A sample is considered a core point if its $\epsilon$-neighborhood contains at least 3 samples; samples that cannot be assigned to any cluster are labeled as noise.

\subsection{Affinity Propagation}

Affinity Propagation is a message-passing clustering algorithm based on pairwise similarities between samples. It treats every sample as a potential exemplar and iteratively exchanges responsibility and availability messages between samples. After convergence, a subset of samples is selected as exemplars, and other samples are assigned to their corresponding exemplars. Therefore, Affinity Propagation does not require the number of clusters to be specified beforehand.

In our experiments, Affinity Propagation uses the standard similarity formulation provided by the implementation. The damping factor \texttt{damping} is set to 0.9 to improve the stability of message passing. This method does not use the HAC distance threshold $\tau$; its clustering result is determined by the pairwise similarity structure and the iterative convergence process.

\subsection{Mean Shift}

Mean Shift is a mode-seeking clustering algorithm based on kernel density estimation. Starting from the sample points, the algorithm iteratively shifts each point toward regions with higher estimated density in the embedding space. Samples that converge to the same density mode are assigned to the same cluster. Similar to DBSCAN, Mean Shift does not require the number of clusters to be specified in advance.

The main parameter of Mean Shift is the bandwidth, which controls the neighborhood scale used for density estimation. In our experiments, the bandwidth is estimated separately for each query from its document embeddings using the \texttt{estimate\_bandwidth} function with \texttt{quantile=0.2}. Since each query only contains a limited number of candidate documents, this setting provides an adaptive estimate of the local density scale for the current sample set.

\section{Prompt Templates}
\newtcolorbox{promptbox}[1]{
    colback=white,
    colframe=black,
    fonttitle=\bfseries,
    title=#1,
    arc=0mm,
    boxrule=0.8pt
}

\begin{promptbox}{Final Answer Generation}
System: You are a helpful assistant, below is a query from a user and some relevant contexts. \
Answer the question given the information in those contexts. Your answer should be short and concise. \
If you cannot find the answer to the question, just say "I don't know".
\\
\textbf{Contexts:} [Filtered Documents] \\
\textbf{Question:} [Question] \\
\end{promptbox}

\section{The Use of LLM}

Large Language Models (LLMs) were used in this work primarily for language polishing and writing assistance. Specifically, LLM-based tools were employed to improve grammar, readability, fluency, and overall clarity of the manuscript.

The core research ideas, methodology design, experiments, analysis, and conclusions were developed and verified entirely by the authors.

\end{document}